\documentclass[]{article}

\usepackage[affil-it]{authblk}
\usepackage{a4wide}
\usepackage{amsmath}
\usepackage{multirow}
\usepackage{booktabs}
\usepackage[ruled,vlined]{algorithm2e}
\usepackage{array}
\usepackage{color}
\usepackage[normalem]{ulem}
\usepackage{hyperref}
\usepackage{amssymb}
\usepackage{graphicx}
\usepackage{amsfonts}
\usepackage[english]{babel}
\usepackage[sort&compress,square,comma,authoryear]{natbib}

\begin{document}
\bibliographystyle{apalike}

\title{CLEAN algorithm implementation comparisons between popular software packages}

\author[1]{Daniel~Wright}
\author[1]{Karel~Ad\'{a}mek}
\author[1]{Wesley~Armour \thanks{E-mail address: \texttt{wes.armour@oerc.ox.ac.uk}} } 

\affil{Oxford e-Research Centre, Department of Engineering Sciences, University of Oxford, 7 Keble road, OX1 3QG, Oxford, United Kingdom}

\maketitle

\begin{abstract}
The CLEAN algorithm, first published by \citet{hogbom74} and its later variants such as Multiscale CLEAN (msCLEAN) by \citet{cornwell08}, has been the most popular tool for deconvolution in radio astronomy. Interferometric imaging used in aperture synthesis radio telescopes requires deconvolution for removal of the telescopes point spread function from the observed images. We have compared source fluxes produced by different implementations of H\"{o}gbom and msCLEAN (WSCLEAN (\citep{offringa-wsclean-2014}, \citep{offringa-wsclean-2017}), CASA \citep{The_CASA_Team_2022}) with a prototype implementation of H\"{o}gbom and msCLEAN for the Square Kilometer Array (SKA) on two datasets. First is a simulation of multiple point sources of known intensity using H\"{o}gbom, where none of the software packages detected all the simulated point sources to within 1.0\% of the simulated values. The second is of supernova remnant G055.7+3.4 taken by the Karl G. Jansky Very Large Array (VLA) using msCLEAN, where each of the software packages produced different images for the same settings.

\end{abstract}

\section{Introduction}
Synthesis imaging is a fundamental technique used in radio astronomy to compute the sky brightness distribution (an image) from frequency elements (visibilities) measured by the radio interferometer. As the visibilities are being sampled in the frequency domain by a sampling function, the resulting image produced by the inverse Fourier transform is convolved with the point spread function (PSF). In the case of radio interferometry, the sampling function is very sparse as a radio interferometer has a limited number of antennas that observe the sky, making the PSF present in the image particularly obstructive to science. The actual image $I_\mathrm{Actual}$ is therefore convolved with PSF giving us so-called dirty image $I_\mathrm{Dirty}$:

\begin{displaymath}
I_\mathrm{Dirty} = I_\mathrm{Actual} * \mathrm{PSF}\,.
\end{displaymath}

In order to recover the actual image, the PSF must be deconvolved from the dirty image. Due to the presence of noise in the image, it is not possible to use deconvolution directly, so a number of algorithms have been developed to overcome this. One of these is the CLEAN algorithm.

\subsection{CLEAN Algorithms}
The CLEAN algorithm, as first published by~\citet{hogbom74} and in its further versions, is a matching pursuit non-linear iterative deconvolution algorithm. In this work, we demonstrate an implementation of two variants of CLEAN:

H\"{o}gbom CLEAN~\citep{hogbom74}: Iteratively scans the image for a maxima, each time one is found, the PSF is centred on the point and a fraction of it is subtracted from the image. A fraction of the maximum value is added to a model image at the same position as the maxima was detected. This continues until an iteration limit or the maxima found is below a set threshold. Finally, the model image is convolved with an idealised telescope beam and any residual flux left in the original image is added to the model image to produce the deconvolved image.

msCLEAN~\citep{cornwell08}: Uses the same principles as H\"{o}gbom, but to better fit the extended emission structures it uses scaling kernels to produce scaled versions of the image. The maxima is searched for across all the scaled versions, with a bias value used to enable comparisons of maximum values across different scales. PSFs are also scaled with the same kernels, to allow PSF subtraction from each scaled image.

H\"{o}gbom CLEAN is more suited to point source images, and msCLEAN is more suited to extended emissions. That is how we deploy and test these algorithms in this work.

\begin{figure*}[htb]
	\centering
	\includegraphics[width=.55\linewidth]{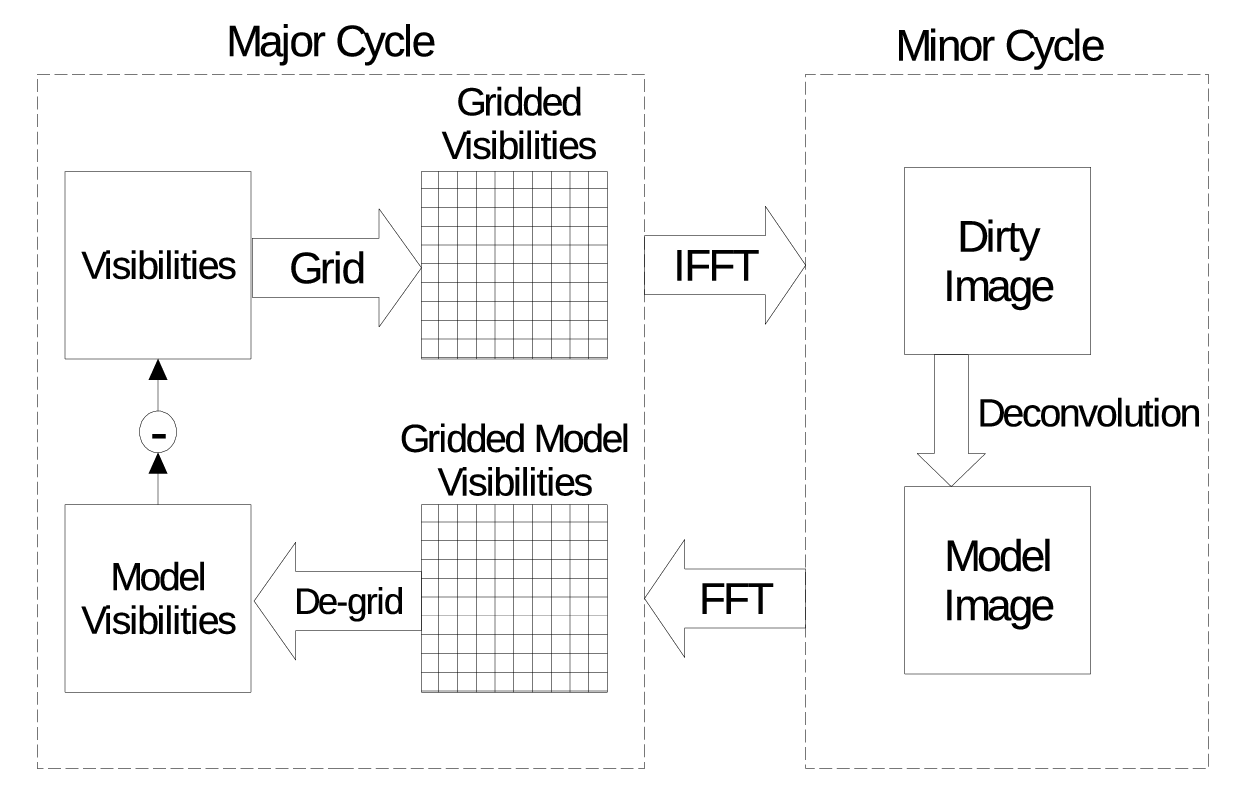}
	\caption{\label{maj_min} Diagram of the Major/Minor Cycle of Image Reconstruction used in radio iterferometry}
\end{figure*}

Image reconstruction in radio interferometry is often performed in a major/minor cycle loop, as is shown in the Figure~\ref{maj_min}. This work is focused on the performance of the minor cycle deconvolution only, in order to assess the previously described CLEAN algorithms, no major cycles have been performed on the test data. 

In this work we use existing packages for deconvolution of radio images (CASA, WSCLEAN) along with a prototype Square Kilometer Array (SKA) software package to produce the output of deconvolution.

\subsection{Datasets}

\begin{figure*}[htb]
     \centering
     \includegraphics[width=.65\linewidth]{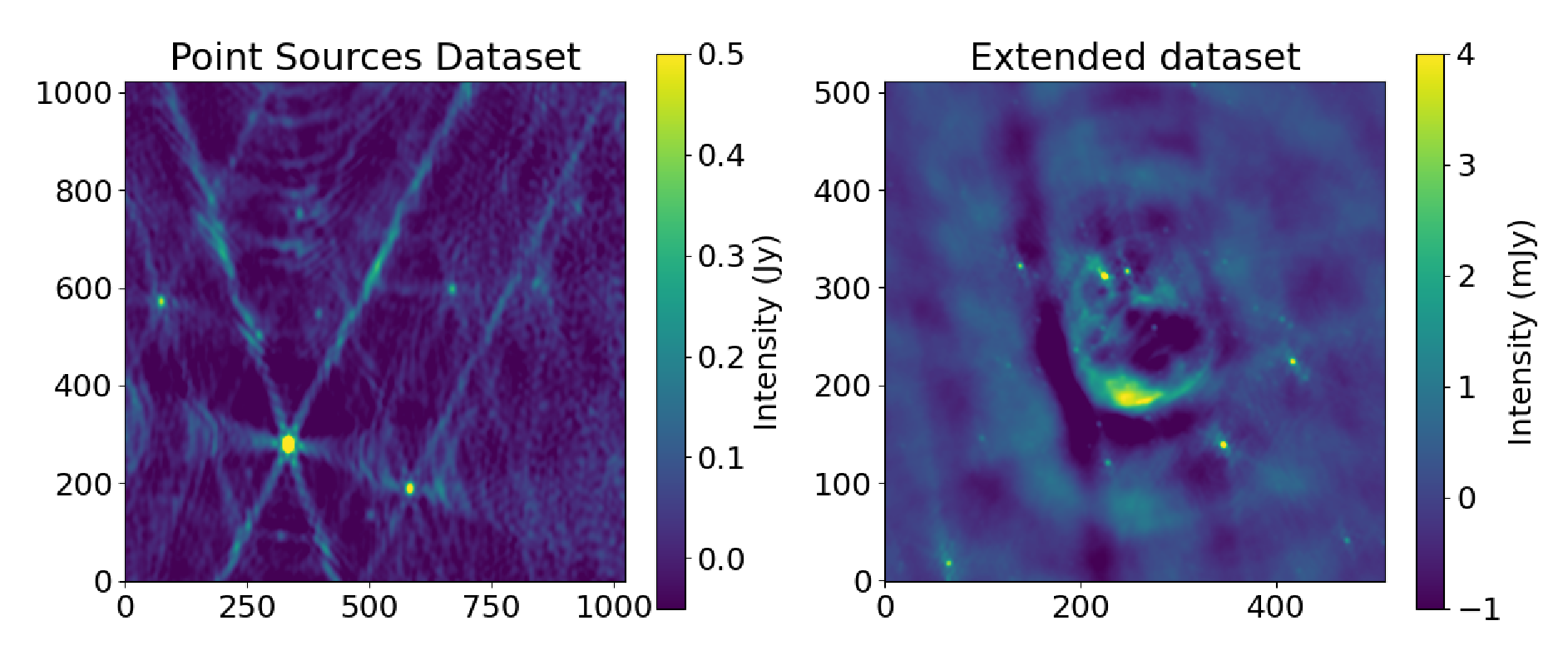}
     \caption{\label{orig_both} Dirty images of point (left) and extended emission (right) sources.}
 \end{figure*}

The simulated point source dataset is a dirty image which contains 30 sources of various locations and intensities. The simulated instrument is the VLA in configuration D. The simulation was performed with the OSKAR software~\citep{dulwich_2020_3758491}. This dirty image is shown in Figure~\ref{orig_both} on the left. The extended emission dataset is a dirty image from an observation of the supernova remnant G055.7+3.4 taken by the VLA in configuration D on 23rd August 2010 \citep{data}.  The dirty image is shown in the Figure~\ref{orig_both} on the right.

\section{Results}
\begin{figure*}[htb]
     \centering
     \includegraphics[width=1\linewidth]{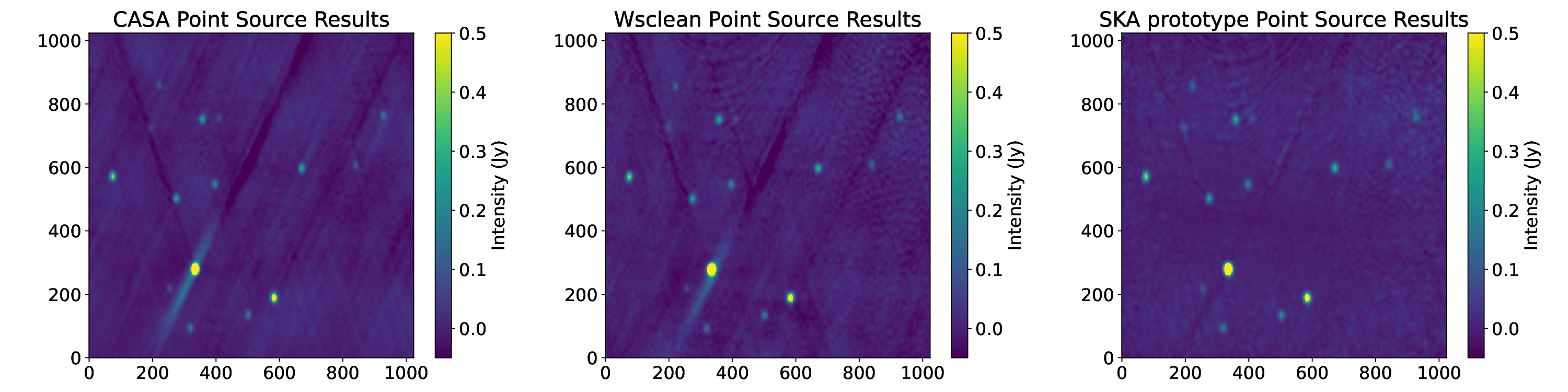}
     \caption{\label{hogbom_res} H\"{o}gbom CLEAN deconvolved images.}
 \end{figure*}

The parameters controlling threshold, gain per cycle and cycle limit were kept consistent across the different packages (CASA, WSCLEAN, SKA prototype) used for all tests. H\"{o}gbom CLEAN was run in each package with a maximum iteration limit of 3,000 and a minimum flux threshold of 1mJy. These results are shown in Figure~\ref{hogbom_res}.

\begin{figure*}[htb]
     \centering
     \includegraphics[width=1\textwidth]{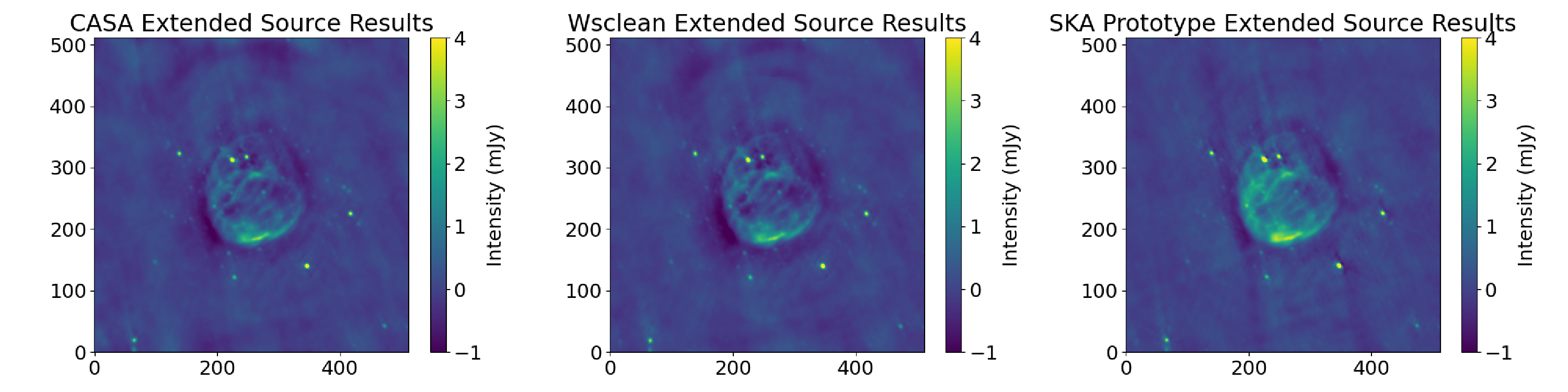}
     \caption{\label{ms_res}msCLEAN deconvolved images.}
 \end{figure*}

The msCLEAN implementations were run with a limit of 1,000 iterations and a threshold of 0.12mJy. Large differences can be seen in the flux and structure of the images produced. These results are shown in the Figure~\ref{ms_res}.

\section{Conclusion}
For the H\"{o}gbom results, the flux of each source deconvolved is measured and its flux compared to the known flux input to the starting simulation, any deviation is noted.

\begin{itemize}

\item WSCLEAN: mean = -3.45\%, stdev = 11.31\%, min = -29.37\%, max = 19.20\%
\item CASA: mean = -5.17\%, stdev = 12.23\%, min = -36.16\%, max = 11.18\%
\item SKA prototype: mean = -1.01\%, stdev = 6.81\%, min = -7.21\%, max = 16.91\%
\end{itemize}
It can be concluded that none of the H\"{o}gbom implementations are particularly accurate without the aid of a major cycle. 
It can be concluded that the SKA prototype is well-aligned with other packages.

\begin{figure*}[htb]
     \centering
     \includegraphics[width=.45\textwidth]{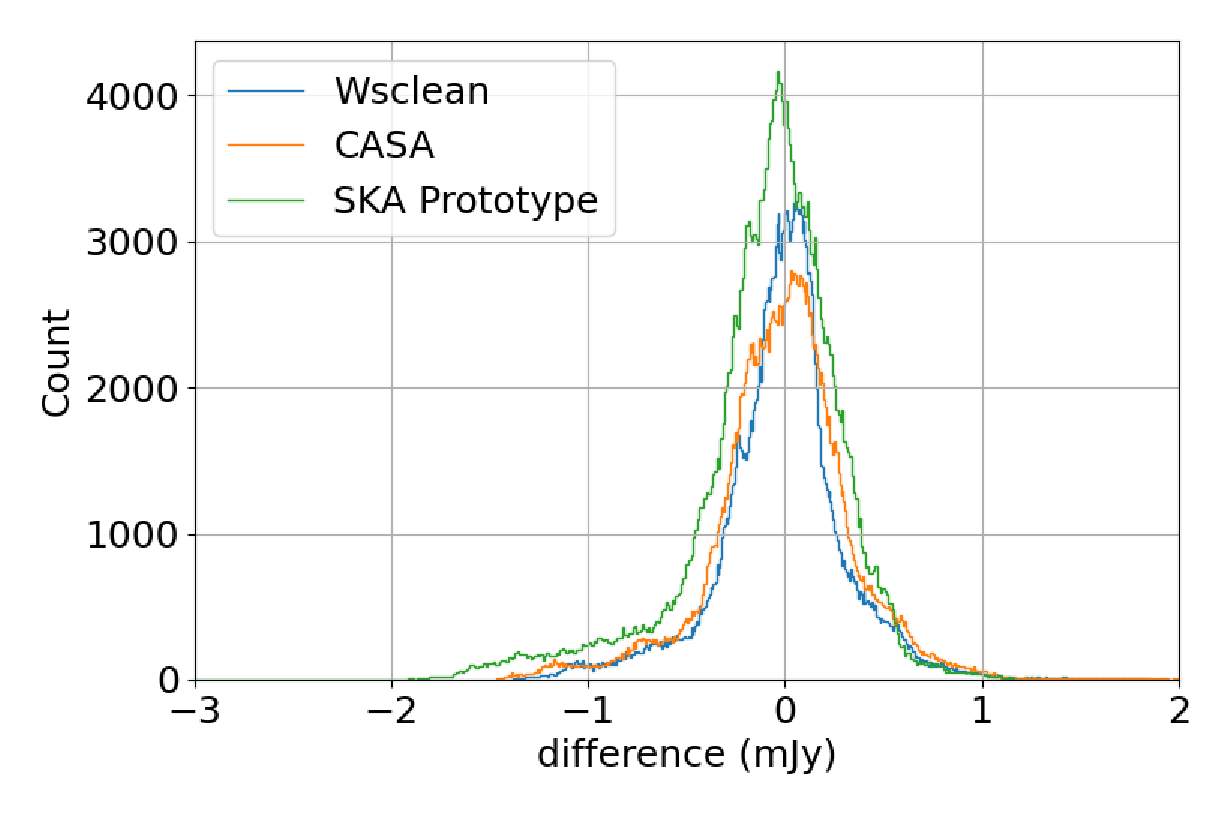}
     \caption{\label{hist_diff}Difference Between Input Dirty Image and CLEANed Output Results}
 \end{figure*}

The msCLEAN results are compared back to the dirty image to find the difference and therefore the change in flux in the CLEANed image, this is shown in the Figure~\ref{hist_diff}.

Figure~\ref{hist_diff} shows that different amounts of flux have been removed by each implementation. The differences in the results of the algorithms show that the various optimisations and design choices used in each of the packages have a large effect on the final image, they do not produce matching images, even though they are using the same underlying algorithm.
The SKA prototype of msCLEAN has the correct shape but performs deeper CLEAN than other packages.

\bibliography{P303_arxiv}  

\section*{Acknowledgements} This work was supported by the Science and Technology Facilities Council [grant number ST/W001969/1].

The National Radio Astronomy Observatory is a facility of the National Science Foundation operated under cooperative agreement by Associated Universities, Inc.

\end{document}